\documentclass{ws-ijmpa}
\usepackage[super,compress]{cite}
\usepackage{graphicx}

\begin{document}
\markboth{Y. Yang et al}{The generalized KG oscillator with PDM in a particular G\"{o}del-type space-time}

%
\catchline{}{}{}{}{}
%

\title{The generalized Klein-Gordon oscillator with position-dependent mass in a particular G\"{o}del-type space-time}

\author{Yi Yang$^1$\footnote{gs.yangyi17@gzu.edu.cn}, Zheng-Wen Long$^1$\footnote{zwlong@gzu.edu.cn (Corresponding author)}, Qi-Kang Ran$^2$, Hao Chen$^1$, Zi-Long Zhao$^1$ and Chao-Yun Long$^1$
}

\address{$^1$College of Physics, Guizhou University, Guiyang, 550025, China\\
$^2$College of Mathematics, Shanghai University of Finance and Economics, Shanghai 200433, China}

\maketitle

\begin{history}
\end{history}

\begin{abstract}
The relativistic quantum dynamics of the generalized Klein-Gordon (KG) oscillator having position-dependent mass in the G\"{o}del-type space-time is investigated. We have presented the generalized KG oscillator in this space-time, and discussed the effect of Cornell potential and linear potential for our considered system. The modification from the parameters of position-dependent mass and characterizing the space-time for the energy spectrums are presented.

\end{abstract}

\ccode{PACS numbers: 03.65.Pm, 03.65.Ge, 98.80.Cq, 11.27.+d}


\section{Introduction}	
The G\"{o}del metric \cite{go1949} is a vital solution in general relativity, which is the first cosmological solution having rotating matter. The metric is characterized by being stable, cylindrically symmetric and spatially homogeneous. In this metric, the causality is broken denoting the existing of the closed time-like curve, and the causality was discussed in detail by these authors \cite{mr1985,mo1988}. The G\"{o}del-type space-time has attracted widespread attention since it was proposed. The authors Figueiredo et al. have investigated the quantum dynamics of Klein-Gordon and Dirac particles in G\"{o}del-type space-time \cite{figue}. Recently, some authors have studied the dynamics of scalar particles and spin-1/2 particles in G\"{o}del-type space-time. In Ref. \cite{jc2016,drukker,das}, the authors studied the influence of topological defects on scalar particles in Som-Raychaudhuri space-time (a special case of the G\"{o}del-type space-time). In Som-Raychaudhuri space-time, the effect of linear scalar potential on scalar particles in an existing topological defect was studied by Bakke et al. \cite{rll7844}. The dynamics of scalar particles in the G\"{o}del-type space-time under an existing external field was investigated \cite{fa588}. The behavior of spin-1/2 particles under this space-time was investigated \cite{fa534}. Moreover, numerous valuable works about the G\"{o}del-type space-time can be found in Refs. \cite{cf047702,ja96,hh83,lb53,ps541,eo530,mm47,kk123}.

As we all know, Moshinsky and Szczepaniak introduced the Dirac oscillator as the famous model of the harmonic oscillator \cite{mo817}. Bruce and Minning introduced KG oscillator \cite{bs711} according to the Dirac oscillator by considering a coupling to the KG equation. The KG oscillator has been studied in various fields, for example PT-symmetric Hamiltonian \cite{pt}, non-commutative space \cite{liang1250047} and curved space-time \cite{lcn13}. Especially, the Klein-Gordon oscillator in curved space-time has been extensively studied. The quantum effects on the KG oscillator in the G\"{o}del-type space-time were discussed \cite{fa598}. In Ref. \cite{rll490}, they investigated the Aharonov-Bohm effect for the KG oscillator in the space-time having a screw dislocation. On the other hand, the generalized KG oscillator also obtains numerous attentions. The so called generalized KG oscillator is considered by the way that the momentum
operator becomes $\vec{p} \rightarrow \vec{p}-i m \omega \beta \vec{r}$, where the coordinate $r$ is substituted
by the potential function $f(r)$. In our previous works, the influence of  topological defect on the generalized Kemmer oscillator under the cosmic string space-time background was discussed \cite{mycpb}, and the generalized Duffin-Kemmer-Petiau (DKP) oscillator under the chiral conical space-time was studied \cite{ch2050107}.

The solutions of KG and Dirac equations have been used to study some problems. The theoretical physicists have developed many quantum mechanics techniques to study the relativistic wave equations, which involve the Nikiforov-Uvarov (NU) method \cite{nu1988,nu080301,hh022104,hh541,mc704,mg123}, the functional Bethe ansatz method \cite{zhang015006,zhang246}, the biconfluent Heun equation method \cite{heun1995,fa211,fa108,fa7943436,rll36} etc.. In this work, we will use the biconfluent Heun equation method to study the generalized KG oscillator subjected to the Cornell potential under the G\"{o}del-type space-time. As we all know, the position-dependent mass (PDM) plays an significant role in relativistic quantum systems. The quantum systems with the PDM have attracted the attentions of physicists. In Ref. \cite{rll161}, the effect of PDM for scalar particle in a dislocation space-time was investigated. In the situation with the uniform magnetic field, position-dependent mass scalar particle was studied \cite{6740360}. Therefore, in this work we will consider the PDM into the generalized KG oscillator under the above space-time to explore the effect of PDM for our considered system.

The structure of this paper is as follows. In the next section, the generalized KG oscillator under the G\"{o}del-type space-time is introduced. In section \ref{sec:3}, the generalized KG oscillator subjected to the Cornell potential is studied under this space-time by using the biconfluent Heun equation method. In section \ref{sec:4}, the effect of the PDM for the generalized KG oscillator is investigated. Finally, the conclusions are given in section \ref{sec:summary}.
\section{Generalized KG oscillator under a particular G\"{o}del-type space-time}\label{sec:two}
The G\"{o}del-type space-time, which does not include horizons and singularities, cannot fully describe real situation of our universe. But it can be used to investigate a number of physical and mathematical models about gravitational backgrounds with rotation and violation of causality in general relativity. In our work, we consider the following line elements of particular G\"{o}del-type space-time background in this coordinates $\left(x^{0}=t, x^{1}=x,\\ x^{2}=y, x^{3}=z\right)$ \cite{fa534,fa598,fa735,fa104,fa2050031,fa025103}
\begin{equation}\begin{aligned}\label{linem}
d s^{2} &=-d t^{2}+d x^{2}+\Big(1-\alpha^{2} x^{2}\Big) d y^{2}-2 \alpha x d t d y+d z^{2} \\
&=-\Big[d t+H(x) d y\Big]^{2}+d x^{2}+D^{2}(x) d y^{2}+d z^{2},
\end{aligned}\end{equation}
where $\alpha > 0$ is real, and the function $H(x)=\alpha x$
and $D(x)=1$. The ranges of the coordinates ($t, x, y, z$) are from $-\infty$ to $\infty$. We can write the line element of this space-time (\ref{linem}) in the form of the covariant metric tensor

\begin{equation}g_{\mu \nu}=\left(\begin{array}{cccc}
-1 & 0 & -\alpha x & 0 \\
0 & 1 & 0 & 0 \\
-\alpha x \quad& 0 & \quad1-\alpha ^2x^2 \quad& 0 \\
0 & 0 & 0 & 1
\end{array}\right),\end{equation}
and its contravariant metric tensor
\begin{equation}g^{\mu \nu}=\left(\begin{array}{cccc}
\alpha ^2x^2-1 \quad& 0 & \quad-\alpha x \quad& 0 \\
0 & 1 & 0 & 0 \\
-\alpha x & 0 & 1 & 0 \\
0 & 0 & 0 & 1
\end{array}\right).\end{equation}
According to the covariant metric tensor, it's determinant can be obtained $detg =-1$, which indicates that the metric of this space-time has the characteristic of being regular everywhere.

In relativistic quantum dynamics, the KG equation describe the particles of spin-0, which can be written as \cite{lcn186,erf2051}
\begin{equation}\frac{1}{\sqrt{-g}} \partial_{\mu}\left(\sqrt{-g} g^{\mu \nu} \partial_{\nu} \Psi\right)=m^{2} \Psi ,\end{equation}
where $g$ and $\partial_{\nu}$ denotes the determinant of metric tensor and the ordinary derivative, respectively.
The KG oscillator is usually considered by the momentum operator
\begin{equation}p_{\mu} \rightarrow p_{\mu}+i m \Omega X_{\mu},
\end{equation}
where $\Omega$ represents the oscillator's frequency and $X_\mu = (0, r , 0,0)$. The generalized KG oscillator is introduced by the substitution $X_\mu= (0, r ,0,0)\rightarrow X_\mu= (0, f (r ),0,0)$. So, we can obtain the generalized KG oscillator
\begin{equation}\begin{aligned}
\frac{1}{\sqrt{-g}}\Big(\partial_{\mu}+m \Omega X_\mu\Big) \sqrt{-g} g^{\mu \nu}\Big(\partial_{\nu}-m \Omega X_{\nu}\Big) \Psi -m^{2} \Psi=0.
\end{aligned}\end{equation}
According to the line element of particular G\"{o}del-type space-time (\ref{linem}), following differential equation can be derived
\begin{equation}\begin{aligned}\label{eq7}
&{\left[-\frac{\partial^{2}}{\partial t^{2}}+\left(\alpha x \frac{\partial}{\partial t}-\frac{\partial}{\partial y}\right)^{2}+\frac{\partial^{2}}{\partial z^{2}}\right] \Psi}
+\left(\frac{\partial}{\partial x}+m \Omega f(x)\right)\left(\frac{\partial \Psi}{\partial x}-m \Omega f(x) \Psi\right)\\
&-m^{2} \Psi=0.
\end{aligned}\end{equation}
The variables ($t,x,y,z$) are independent, therefore we can select the ansatz
\begin{equation}\label{ansatz}
\Psi(t, x, y, z)=e^{i\left(-E t+l y+k z\right)} \psi(x),
\end{equation}
where $E$ is energy. According to Eq. (\ref{eq7}) and Eq. (\ref{ansatz}), we can obtain
\begin{equation}\label{eq9}
\frac{d^2\psi(x)}{dx^2}-\bigg[\alpha^{2} E^{2} x^{2}+m^{2} \Omega^{2} f^{2}(x)+2 \alpha E l x\bigg] \psi(x)=\bigg[m^2+l^2+k^2-E^2+m\Omega \frac{df(x)}{dx} \bigg]\psi(x).
\end{equation}
In the next section, we will study the Eq. (\ref{eq9}) through considering the function $f(x)=A x+\frac{B}{x}$.

\section{Generalized KG oscillator having the Cornell potential}\label{sec:3}
The Cornell potential has an important effect on the study of mesonic, bound heavy quarks, and antiquark in particle physics systems \cite{eichten,cp167}. It has been successfully used to describe experimental data in particle physics \cite{perkin}. We consider the generalized KG oscillator having the Cornell potential to observe the effect of Cornell potential on the energy spectrum of a spin-0 scalar particle. The Cornell potential can be read as \cite{cnr596,ch2050179,1950314,1248393}
\begin{equation}
f(x)=A x+\frac{B}{x},
\end{equation}
where the Cornell potential includes the linear term and Coulomb term. The linear term is a confined term, which can represent the non-perturbative effects of quantum chromodynamics. The second term is the non-confining term, which comes from the one-gluon exchange between the quark and antiquark.
Substituting the Cornell potential into Eq. (\ref{eq9}), which becomes
\begin{equation}
\frac{d^2\psi(x)}{dx^2}-\bigg[\omega^{2} x^{2}+2 \alpha E l x+2ABm^{2} \Omega^{2} +\frac{m^{2} \Omega^{2}B^2-m\Omega B}{x^2}\bigg] \psi(x)=\beta_0\psi(x),
\end{equation}
where
\begin{equation}
\omega=\sqrt{\alpha^{2} E^{2}+m^{2} \Omega^{2}A^2},
~\beta_0=\left(m^{2}+l^{2}+k^{2}-E^{2}+m\Omega A\right).
\end{equation}
We take the define $r=\sqrt{\omega} x$, then the above equation becomes
\begin{equation}\label{eq13}
\frac{d^2\psi(r)}{dr^2}-\left[r^{2}+\Lambda_1 r+\frac{2ABm^{2} \Omega^{2}}{\omega} +\frac{m^{2} \Omega^{2}B^2-m\Omega B}{r^2}\right] \psi(x)=\Lambda_2\psi(x),
\end{equation}
where
\begin{equation}
\Lambda_1=\frac{2 \alpha E l}{\omega^{3 / 2}}, ~ \Lambda_2=\frac{\beta_{0}}{\omega}.
\end{equation}

Let the solution of Eq. (\ref{eq13}) is
\begin{equation}\label{eq15}
\psi(r)=r^{\eta} exp\left[-\frac{1}{2}\Big(\Lambda_1r+r^2\Big)\right] H(r), \end{equation}
where
\begin{equation}
\eta=\frac{1}{2}\left[1+\sqrt{1+4 \Big(m^{2} \Omega^{2}B^2-m\Omega B\Big)}\right].
\end{equation}
Substituting Eq. (\ref{eq15}) into Eq. (\ref{eq13}), we obtain
\begin{equation}\begin{aligned}\label{eq17}
r \frac{d^2H(r)}{dr^2}+\Big[2 \eta-\Lambda_1 r-2 r^{2}\Big] \frac{dH(r)}{dr}
+\Big[\left(\lambda-\xi-2\right)r-\Lambda_1\eta\Big] H(r)=0,
\end{aligned}\end{equation}
where $\lambda$ and $\xi$ are given by
\begin{align}
\lambda&=\frac{\Lambda_1^2}{4}-\frac{2ABm^{2}\Omega^{2}+\beta_0}{\omega},\\
\xi&=\sqrt{1+4 \Big(m^{2} \Omega^{2}B^2-m\Omega B\Big)}.
\end{align}
One can see that the Eq. (\ref{eq17}) has the same mathematical expression as the biconfluent Heun equation \cite{er2051,kk86,kk2838,sh240,hm1950054}. The biconflent Heun $H(r)$ is given by \cite{er2051}
\begin{equation}
H(r)=H_{B}\Big(2\eta-1, \Lambda_1, \lambda, 0; r\Big).
\end{equation}
Using the Frobenius method \cite{fc90}, we can obtain the bound states solutions. Therefore, a power series expansion of the solution for the Eq. (\ref{eq17}) in the origin can be given by
\begin{equation}\label{eq21}
H(r)=\sum_{j=0}^{\infty} A_{j} r^{j}.
\end{equation}
Using the Eq. (\ref{eq21}) and the Eq. (\ref{eq17}), the recurrence relation for the coefficients can be written as
\begin{align}
A_{n+2}=& \frac{1}{(n+2)(n+2 \eta+1)}\bigg[\Lambda_1\Big(n+\eta+1\Big)A_{n+1}
+\Big(2 n-\lambda+2\eta+1\Big) A_{n}\bigg], \quad n \geq 1.
\end{align}
The coefficients $A_1$ and $A_2$ are
\begin{equation}\begin{aligned}
A_{1}=&\frac{\Lambda_1}{2}A_{0}=\frac{ \alpha E l}{\omega^{3/2}} A_{0}, \\
A_{2}=& \frac{1}{2(2 \eta+1)}\left[\Lambda_1\Big(\eta+1\Big)A_{1}
+\Big(1-\lambda-2\eta\Big) A_{0}\right].
\end{aligned}\end{equation}
In order to make the function become a polynomial of $r^n$, the two conditions are satisfied
\begin{align}
\lambda-\xi-2&=2 n, \quad n=0,1,2 \cdots, \label{fc}\\
A_{n+1}&=0,
\end{align}
where the polynomial $A_{n+1}$ has $n+1$ real roots when $1+\xi > 0$ and $\Lambda_1\in R$, which can be
written as a determinant of dimension $n + 1$

\begin{equation}\begin{array}{|ccccccc|}
-\Lambda_1\eta & 1 & 0 & 0 & \cdots & \cdots & 0 \\
4n\eta & -\Lambda_1\eta -\Lambda_1 & 1 & 0 & \cdots & \cdots & 0 \\
0 & 4\big(1+2\eta\big)\big(n-1\big) &  -\Lambda_1\eta -2\Lambda_1 & 1 & 0 &\cdots & 0 \\
0 & 0 & \Sigma_{2} & -\Lambda_1\eta -3\Lambda_1 & 1 & \cdots & 0 \\
\vdots & \vdots & 0 & \ddots & \ddots & \ddots & \cdots \\
\vdots & \vdots & \vdots & \vdots & \Sigma_{j-1} & \prod_{j-1} & 1 \\
0 & 0 & 0 & 0 & 0 & \Sigma_{j} & \prod_{j}
\end{array}=0,\end{equation}
where $\prod_{j}=-\Lambda_1\eta -\Lambda_1(j+1)$ and $\Sigma_{j}=2(j+1)(j+2\eta)(n-j)$.

From the Eq. (\ref{fc}) the energy spectrum is given by
\begin{equation}
\begin{aligned}
&\frac{\alpha^{2} l^{2}E^{2}}{\omega^{2}}-\Big(m^{2}+l^{2}+k^{2}-E^{2}+m\Omega A+2ABm^2\Omega^2\Big)\\
&-\Bigg[2+\sqrt{1+4 \Big(m^{2} \Omega^{2}B^2-m\Omega B\Big)}\Bigg]\omega=2 n \omega,
\end{aligned}
\end{equation}
and the wave function is given by
\begin{equation}
\begin{aligned}
\psi(x)&=\Big(\sqrt{\omega}x\Big)^{\eta} exp\left[-\frac{1}{2}\Big(\frac{2 \alpha E l}{\omega}x+\omega x^2\Big)\right]\\
&\times H_{B}\Big(\sqrt{1+4 \Big(m^{2} \Omega^{2}B^2-m\Omega B\Big)}, \Lambda_1, \lambda, 0; \sqrt{\omega}x\Big).
\end{aligned}
\end{equation}

\section{Generalized KG oscillator with the position-dependent mass}\label{sec:4}
In this section, we will study the effect of the position-dependent mass for the generalized KG oscillator in our considered space-time background. The position-dependent mass is given by \cite{aa015006,tq015007,bq1850025}
\begin{equation}
m(x)=m_{0}\left(1+\frac{\mathbb{K} \mathbb{C}}{x}\right),
\end{equation}
where $m_0$, $\mathbb{C}$ and $\mathbb{K}=\frac{\hbar}{M_{0} c}$
are the rest mass of the particle, the
real constant and the Compton-like wavelength, respectively.

When we consider the above position-dependent mass, the equation of the KG oscillator becomes
\begin{equation}\begin{aligned}
&\frac{1}{\sqrt{-g}}\bigg[\partial_{\mu}+m_{0}\left(1+\frac{\mathbb{K} \mathbb{C}}{x}\right) \Omega X_\mu\bigg] \sqrt{-g} g^{\mu \nu}\bigg[\partial_{\nu}-m_{0}\left(1+\frac{\mathbb{K} \mathbb{C}}{x}\right) \Omega X_{\nu}\bigg] \Psi \\
&=m^2_{0}\left(1+\frac{\mathbb{K} \mathbb{C}}{x}\right)^{2} \Psi.
\end{aligned}\end{equation}
According to the line element (\ref{linem}), the differential equation is given by
\begin{equation}\begin{aligned}\label{eq31}
&{\bigg[-\frac{\partial^{2}}{\partial t^{2}}+\left(\alpha x \frac{\partial}{\partial t}-\frac{\partial}{\partial y}\right)^{2}+\frac{\partial^{2}}{\partial z^{2}}\bigg] \Psi}
+\bigg[\frac{\partial}{\partial x}+m_{0}\left(1+\frac{\mathbb{K} \mathbb{C}}{x}\right) \Omega f(x)\bigg]\\
&\times\bigg[\frac{\partial \Psi}{\partial x}-m_{0}\left(1+\frac{\mathbb{K} \mathbb{C}}{x}\right) \Omega f(x) \Psi\bigg]=\bigg[m_{0}\left(1+\frac{\mathbb{K} \mathbb{C}}{x}\right)\bigg]^{2} \Psi.
\end{aligned}\end{equation}
Inserting the ansatz (\ref{ansatz}) into Eq. (\ref{eq31}), we obtain
\begin{equation}\begin{aligned}
&\frac{d^2\psi(x)}{dx^2}-\bigg[\alpha^{2} E^{2} x^{2}+2 \alpha E l x+m^{2}_0\left(1+\frac{\mathbb{K} \mathbb{C}}{x}\right)^2 \Omega^{2} f^{2}(x)+\frac{m_0\Omega\mathbb{K} \mathbb{C}f(x)}{x^2}\\
&-m_{0}\left(1+\frac{\mathbb{K} \mathbb{C}}{x}\right)\Omega \frac{df(x)}{dx} \bigg] \psi(x)=\bigg[l^2+k^2-E^2+m^{2}_0\left(1+\frac{\mathbb{K} \mathbb{C}}{x}\right)^2 \bigg]\psi(x).
\end{aligned}\end{equation}

The linear scalar potential has been extensively studied in atomic, molecular physics \cite{killing,fonse}, and relativistic quantum mechanics \cite{plante,tezuka,vitoria370}. Meanwhile, the linear confinement of quantum particles is of significance to the quark confinement model \cite{261}. After introducing the position-dependent mass, we study the linear confinement of a spin-0 particle under the G\"{o}del-type space-time. The linear potential can be given by \cite{1550002,jy378,sm122108,hh423,hh566,wg2000}
\begin{equation}
f(x)=\Xi x,
\end{equation}
where $\Xi >0$ is a constant. When we consider the linear potential, we can get
\begin{equation}
\begin{aligned}
&\frac{d^2\psi(x)}{dx^2}-\bigg[\tilde{\omega}^{2} x^{2}+2 \left(\alpha E l+\mathbb{K} \mathbb{C}m^2_0\Omega^2\Xi^2 \right)x+m^{2}_0\mathbb{K}^2 \mathbb{C}^2\Xi^2 \Omega^{2} \\
&+\frac{2m^{2}_0\mathbb{K} \mathbb{C}}{x}+\frac{m^{2}_0\mathbb{K}^2 \mathbb{C}^2}{x^2}\bigg] \psi(x)=\tilde{\beta_0}\psi(x),
\end{aligned}
\label{eq33}
\end{equation}
where
\begin{equation}
\tilde{\omega}=\sqrt{\alpha^{2} E^{2}+m^{2} \Omega^{2}\Xi^2},
~\tilde{\beta_0}=\left(m^{2}_0+l^{2}+k^{2}-E^{2}+m_0\Omega \Xi\right).
\end{equation}
When we define $\rho=\sqrt{\tilde{\omega}} x$, the Eq. (\ref{eq33}) can be written as
\begin{equation}\label{eq35}
\frac{d^2\psi(\rho)}{d\rho^2}-\left[\rho^{2}+\Theta_1 \rho+\frac{m^{2}_0\mathbb{K}^2 \mathbb{C}^2\Xi^2 }{\tilde{\omega}} +\frac{2m^{2}_0\mathbb{K} \mathbb{C}}{\rho\sqrt{\tilde{\omega}}}+\frac{m^{2}_0\mathbb{K}^2 \mathbb{C}^2}{\rho^2}\right] \psi(\rho)=\Theta_2\psi(\rho),
\end{equation}
where
\begin{equation}
\Theta_1=\frac{2\left(\alpha E l+\mathbb{K} \mathbb{C}m^2_0\Omega^2\Xi^2 \right)}{\tilde{\omega}^{3 / 2}}, ~ \Theta_2=\frac{\tilde{\beta_0}}{\tilde{\omega}}.
\end{equation}
The solution of Eq. (\ref{eq35}) is
\begin{equation}\label{eq37}
\psi(\rho)=\rho^{\delta} exp\left[-\frac{1}{2}\Big(\Theta_1\rho+\rho^2\Big)\right] H(\rho), \end{equation}
where
\begin{equation}
\delta=\frac{1}{2}\left[1+\sqrt{1+4 \Big(m^{2}_0\mathbb{K}^2 \mathbb{C}^2\Big)}\right]
\end{equation}
Substituting Eq. (\ref{eq37}) into Eq. (\ref{eq35}), we can obtain
\begin{equation}\begin{aligned}\label{eq39}
&\rho \frac{d^2H(\rho)}{d\rho^2}+\bigg[2 \delta-\Theta_1 \rho-2 \rho^{2}\bigg] \frac{dH(\rho)}{d\rho}
+\bigg[\left(\kappa-\zeta-2\right)\rho\\
&-\frac{1}{2}\left(2\Theta_1\delta+\frac{4m^{2}_0\mathbb{K} \mathbb{C}}{\sqrt{\tilde{\omega}}}\right)\bigg] H(\rho)=0,
\end{aligned}\end{equation}
where $\kappa$ and $\zeta$ can be read as
\begin{align}
\kappa&=\frac{\Theta_1^2}{4}-\frac{m^{2}_0\mathbb{K}^2 \mathbb{C}^2\Xi^2 \Omega^{2}+\tilde{\beta_0}}{\tilde{\omega}},\\
\zeta&=\sqrt{1+4 \Big(m^{2}_0\mathbb{K}^2 \mathbb{C}^2\Big)}.
\end{align}
The Eq. (\ref{eq39}) is the same as the Eq. (\ref{eq17}), which have the standard form of Heun's equation. Therefore, when we consider the position-dependent mass under the particular G\"{o}del-type space-time background, the biconflent Heun $H(\rho)$ is given by
\begin{equation}
H(r)=H_{B}\Big(2\delta-1, \Theta_1, \kappa, \frac{4m^{2}_0\mathbb{K} \mathbb{C}}{\sqrt{\tilde{\omega}}}; \rho\Big).
\end{equation}
Similar to the processing method of above section, the solution can be expanded into the power series by the Frobenius method
\begin{equation}\label{eq43}
H(x)=\sum_{i=0}^{\infty} C_{i} x^{i}.
\end{equation}
Inserting the power series solution (\ref{eq43}) into the Eq. (\ref{eq39}), the relation of the coefficients can be got
\begin{equation}
\begin{aligned}
C_{n+2}&=\frac{1}{(n+2)(n+2 \delta+1)}\bigg\{\Big[\frac{2m^{2}_0\mathbb{K} \mathbb{C}}{\sqrt{\tilde{\omega}}}+\Theta_1\Big(n+\delta+1\Big)\Big]C_{n+1}\\
&+\Big(2 n-\kappa+2\delta+1\Big) C_{n}\bigg\}, \quad n \geq 1.
\end{aligned}
\end{equation}
For example,
$C_1$ and $C_2$ are
\begin{align}
C_{1}=&\frac{1}{2}\left(\Theta_1+\frac{2m^{2}_0\mathbb{K} \mathbb{C}}{\sqrt{\tilde{\omega}}\delta} \right)C_{0}, \\
C_{2}=& \frac{1}{2(2\delta+1)}\bigg\{\Big[\frac{2m^{2}_0\mathbb{K} \mathbb{C}}{\sqrt{\tilde{\omega}}}+\Theta_1\Big(\delta+1\Big)\Big]C_{1}
+\Big(1-\kappa+2\delta\Big) C_{0}\bigg\}.
\end{align}
The two conditions are similarly satisfied
\begin{align}
\kappa-\zeta-2&=2 n, \quad n=0,1,2 \cdots, \label{firstc}\\
C_{n+1}&=0.
\end{align}

From the Eq. (\ref{firstc}), the energy eigenvalues
equation for the generalized KG oscillator with the PDM under our considered space-time background can be derived as
\begin{equation}\begin{aligned}\label{eq49}
E^{2}&+\frac{\left(\alpha E l+\mathbb{K} \mathbb{C}m^2_0\Omega^2\Xi^2 \right)^2}{\tilde{\omega}^{2}}-\Big(m^{2}_0+l^{2}+k^{2}+m_0\Omega \Xi+m^{2}_0\mathbb{K}^2 \mathbb{C}^2\Xi^2 \Omega^{2}\Big)\\
&-\Bigg[1+\sqrt{1+4 \Big(m^{2}_0\mathbb{K}^2 \mathbb{C}^2\Big)}\Bigg]\tilde{\omega}=(2 n+1) \tilde{\omega},
\end{aligned}\end{equation}
and the wave function is given by
\begin{equation}\begin{aligned}
\psi(x)&=\Big(\sqrt{\tilde{\omega}}x\Big)^{\delta} exp\left[-\frac{1}{2}\Big(\frac{2\alpha E l+2\mathbb{K} \mathbb{C}m^2_0\Omega^2\Xi^2 }{\tilde{\omega}}x+\omega x^2\Big)\right]\\
&\times H_{B}\Big(2\delta-1, \Theta_1, \kappa, \frac{4m^{2}_0\mathbb{K} \mathbb{C}}{\sqrt{\tilde{\omega}}}; \sqrt{\tilde{\omega}}x\Big).
\end{aligned}\end{equation}

Let $\alpha =0$, this solution can be reduced to Minkowski metric. Therefore, the energy spectrum Eq. (\ref{eq49}) becomes
\begin{equation}\begin{aligned}\label{eq51}
E^{2}=m^{2}_0+l^{2}+k^{2}+m_0\Omega\Xi\Bigg[2n+3+\sqrt{1+4 \Big(m^{2}_0\mathbb{K}^2 \mathbb{C}^2\Big)}\Bigg].
\end{aligned}\end{equation}
We can know that the Eq. (\ref{eq51}) is the energy levels of the generalized KG oscillator subjected to linear potential under the Minkowski metric.
\section{Conclusions} \label{sec:summary}
In our work, we have studied the generalized KG oscillator with position-dependent mass under the particular G\"{o}del-type space-time. The wave functions and energy levels in the existing of Cornell potential are given in this space-time, which rely on the parameters of the space-time and the Cornell potential. Moreover, the results considered the position-dependent mass are obtained in this space-time by using the biconfluent Heun equation. The influence of linear potential on our researched system with PDM is shown. As a special case, the modification of PDM for the energy levels in the Minkowski metric are also obtained.

\section*{Acknowledgments}
This work is supported by the National Natural Science Foundation of China (Grant No. 11465006 and 11565009) and the Major Research Project of innovative Group of Guizhou province (2018-013).


\end{document}